\documentstyle[12pt]{article}

\begin{document}

\baselineskip 6mm
\renewcommand{\thefootnote}{\fnsymbol{footnote}}

\newcommand{\nc}{\newcommand}
\newcommand{\rnc}{\renewcommand}


\rnc{\baselinestretch}{1.24}    
\setlength{\jot}{6pt}       
\rnc{\arraystretch}{1.24}   

\makeatletter
\rnc{\theequation}{\thesection.\arabic{equation}}
\@addtoreset{equation}{section}
\makeatother



\nc{\be}{\begin{equation}}

\nc{\ee}{\end{equation}}

\nc{\bea}{\begin{eqnarray}}

\nc{\eea}{\end{eqnarray}}

\nc{\xx}{\nonumber\\}

\nc{\ct}{\cite}

\nc{\la}{\label}

\nc{\eq}[1]{(\ref{#1})}

\nc{\newcaption}[1]{\centerline{\parbox{6in}{\caption{#1}}}}

\nc{\fig}[3]{

\begin{figure}
\centerline{\epsfxsize=#1\epsfbox{#2.eps}}
\newcaption{#3. \label{#2}}
\end{figure}
}


\def\CA{{\cal A}}
\def\CC{{\cal C}}
\def\CD{{\cal D}}
\def\CE{{\cal E}}
\def\CF{{\cal F}}
\def\CG{{\cal G}}
\def\CH{{\cal H}}
\def\CK{{\cal K}}
\def\CL{{\cal L}}
\def\CM{{\cal M}}
\def\CN{{\cal N}}
\def\CO{{\cal O}}
\def\CP{{\cal P}}
\def\CR{{\cal R}}
\def\CS{{\cal S}}
\def\CU{{\cal U}}
\def\CW{{\cal W}}
\def\CY{{\cal Y}}
\def\Cop{\bbbc}
\def\Zop{\bbbz}
\def\Rop{\bbbr}
\def\Nop{\bbbn}
\def\bbbz {{\sf Z\!\!Z}}
\def\bbbp {{\sf P\!\!|}}
\def\bbbr {{\rm I\!R}}
\def\bbbn {{\rm I\!N}}
\def\RR{R-R }


\def\IR{{\hbox{{\rm I}\kern-.2em\hbox{\rm R}}}}
\def\IB{{\hbox{{\rm I}\kern-.2em\hbox{\rm B}}}}
\def\IN{{\hbox{{\rm I}\kern-.2em\hbox{\rm N}}}}
\def\IC{\,\,{\hbox{{\rm I}\kern-.59em\hbox{\bf C}}}}
\def\IZ{{\hbox{{\rm Z}\kern-.4em\hbox{\rm Z}}}}
\def\IP{{\hbox{{\rm I}\kern-.2em\hbox{\rm P}}}}
\def\IH{{\hbox{{\rm I}\kern-.4em\hbox{\rm H}}}}
\def\ID{{\hbox{{\rm I}\kern-.2em\hbox{\rm D}}}}


\def\a{\alpha}
\def\b{\beta}
\def\ga{\gamma}
\def\d{\delta}
\def\ep{\epsilon}
\def\ph{\phi}
\def\k{\kappa}
\def\l{\lambda}
\def\m{\mu}
\def\n{\nu}
\def\th{\theta}
\def\rh{\rho}
\def\s{\sigma}
\def\t{\tau}
\def\G{\Gamma}
\def\ra{\rangle}
\def\lan{\langle}
\def\da{\dot{a}}
\def\db{\dot{b}}
\def\G{\Gamma}
\def\D{\Delta}
\def\L{\Lambda}
\def\S{\Sigma}
\def\g{\gamma}
\def\e{\varepsilon}
\def\m{\mu}
\def\tr{\mbox{tr}}
\def\n{\nu}
\def\w{\omega}
\def\O{\Omega}
\def\v{\varrho}
\def\vt{\vartheta}
\def\mc{\mathcal}
\def\N{\nabla}
\def\p{\partial}
\def\ra{\rangle}
\def\dg{\dagger}
\def\wt{\widetilde}
\def\V{|~V_{kin}~\ra}
\def\T{\Theta}


\def\half{\frac{1}{2}}
\def\dint#1#2{\int\limits_{#1}^{#2}}
\def\goto{\rightarrow}
\def\para{\parallel}
\def\brac#1{\langle #1 \rangle}
\def\grad{\nabla}
\def\curl{\nabla\times}
\def\div{\nabla\cdot}
\def\p{\partial}
\def\e{\epsilon_0}


\def\Tr{{\rm Tr}\,}
\def\det{{\rm det}}


\def\vare{\varepsilon}
\def\barz{\bar{z}}
\def\barw{\bar{w}}


\def\ad{\dot{\a}}
\def\bd{\dot{\b}}
\def\cd{\dot{\ga}}
\def\dd{\dot{\d}}
\def\8{SO(8)}
\def\so{SO(2)_1}
\def\soo{\widetilde{SO(4)}}
\def\sop{SO(2)_2}
\def\sur{SU(2)_R}
\def\sul{SU(2)_L}
\def\basis{\sul\times\sur\times\so\times\sop}
\def\bc{{\bf C}}
\def\bfz{{\bf Z}}
\def\bz{\bar{z}}
\def\bq{\bar{q}}
\def\bx{\bar{X}}
\def\bp{\bar{P}}
\def\bl{\bar{\lambda}}
\def\bip{\bar{\IP}}
\def\bir{\bar{\IR}}
\def\bif{\bar{f}}
\def\bim{\IP^+}
\def\bin{\IP^-}
\def\bL{\bar{\Lambda}}
\def\Ta{~\T_{\a}~}
\def\Tad{~\T_{\ad}~}
\def\Tba{~\bar{\T}_{\a}~}
\def\Tbad{~\bar{\T}_{\ad}~}
\def\La{~\bL_{\a}~}
\def\Lb{~\bL_{\b}~}
\def\Lc{~\bL_{\ga}~}
\def\Ld{~\bL_{\d}~}
\def\Lad{~\bL_{\ad}~}
\def\Lbd{~\bL_{\bd}~}
\def\Lcd{~\bL_{\cd}~}
\def\Ldd{~\bL_{\dd}~}
\def\lal{~\l_{\a}~}
\def\lb{~\l_{\b}~}
\def\lc{~\l_{\ga}~}
\def\ld{~\l_{\d}~}
\def\lad{~\l_{\ad}~}
\def\lbd{~\l_{\bd}~}
\def\lcd{~\l_{\cd}~}
\def\ldd{~\l_{\dd}~}
\def\lba{~\bl_{\a}~}
\def\lbb{~\bl_{\b}~}
\def\lbc{~\bl_{\ga}~}
\def\lbd{~\bl_{\d}~}
\def\lbad{~\bl_{\ad}~}
\def\lbbd{~\bl_{\bd}~}
\def\lbcd{~\bl_{\cd}~}
\def\lbdd{~\bl_{\dd}~}
\def\qa{~q^-_{(r)\a}~}
\def\qad{~q^-_{(r)\ad}~}
\def\qab{~\bq^-_{(r)\a}~}
\def\qabd{~\bq^-_{(r)\ad}~}
\def\Qa{~Q^-_{(r)\a}~}
\def\Qad{~Q^-_{(r)\ad}~}
\def\Qab{~\bar{Q}^-_{(r)\a}~}
\def\Qabd{~\bar{Q}^-_{(r)\ad}~}
\def\paad{~\IP_{\a\ad}~}
\def\pada{~\IP_{\ad \a}~}
\def\pbaad{~\bip_{\a\ad}~}
\def\pbada{~\bip_{\ad \a}~}
\def\pbbd{~\IP_{\b\bd}~}
\def\pbdb{~\IP_{\bd \b}~}
\def\pbbbd{~\bip_{\b\bd}~}
\def\pbbdb{~\bip_{\bd \b}~}
\def\xaad{~X_{\a\ad}~}
\def\xada{~X_{\ad \a}~}
\def\xbaad{~X_{\a\ad}~}
\def\xbada{~X_{\ad \a}~}
\def\xp{X^+}
\def\xm{X^-}
\def\xt{\tilde{X}}
\def\pt{\tilde{P}}
\def\xpt{\tilde{X}^+}
\def\xmt{\tilde{X}^-}
\def\pp{P^+}
\def\pn{P^-}
\def\ppt{\tilde{P}^+}
\def\pnt{\tilde{P}^-}
\def\dab{~\delta_{\a\b}~}
\def\dac{~\delta_{\a\ga}~}
\def\dbc{~\delta_{\b\ga}~}
\def\dabd{~\delta_{\ad\bd}~}
\def\dacd{~\delta_{\ad\cd}~}
\def\dbcd{~\delta_{\bd\cd}~}
\def\eab{~\ep_{\a\b}~}
\def\eac{~\ep_{\a\ga}~}
\def\ebc{~\ep_{\b\ga}~}
\def\eabd{~\ep_{\ad\bd}~}
\def\eacd{~\ep_{\ad\cd}~}
\def\ebcd{~\ep_{\bd\cd}~}

\begin{titlepage}

\hfill\parbox{5cm} 
{IP/BBSR/2003-06\\hep-th/0303223} \\

\vspace{25mm}

\begin{center}
{\Large \bf D-branes in PP-wave light cone string field theory }

\vspace{15mm}
B. Chandrasekhar\footnote{chandra@iopb.res.in},
Alok Kumar\footnote{kumar@iopb.res.in}
\\[10mm]
{\sl Institute of Physics, Bhubaneswar 751 005, INDIA} \\
\end{center}

\thispagestyle{empty}

\vskip2cm


\centerline{\bf ABSTRACT}
\vskip 4mm
\noindent

The cubic interaction vertex and the dynamical supercharges are 
constructed for open strings ending on D7-branes, 
in light-cone superstring field theory in PP-wave background. 
In this context, we 
write down the symmetry generators in terms of the relevant 
group structure: $SU(2)\times SU(2)\times SO(2)\times SO(2)$, originating
from the eight transverse directions in the PP-wave background and use the
expressions to explicitly construct the vertex at the level of 
stringy zero modes.
The results are further generalized to include all the stringy excitations
as well. 
\\

\vspace{2cm}

\today

\end{titlepage}

\renewcommand{\thefootnote}{\arabic{footnote}}
\setcounter{footnote}{0}

\section{Introduction}

Maldacena's conjecture~\cite{adscft} of the duality between Type-IIB
superstring theory in $AdS_{5}\times S^{5}$ background and 
~${\cal{N}} = 4 $~ super-Yang-Mills theory has gone through several
nontrivial checks at the level
of supergravity. However, in most of the AdS/CFT dualities, it is not 
possible to go beyond the supergravity approximation on the string side,
making  
its usefulness limited. Recently, the authors of~\cite{BMN} elucidated, that
there exists a new scaling limit, in which a particular
subsector of the ~${\cal{N}} = 4$~ super-Yang-Mills theory gets mapped to
Type-IIB superstring theory in a PP-wave background. PP-wave backgrounds can
be obtained by taking a 'Penrose limit'~\cite{Penrose,Guven,Blau} 
of the geometry near a null geodesic in 
$AdS_5 \times S^5$ carrying a large angular momentum on the $S^5$. 
The metric is then given by,
\be \la{ppwave}
ds^2 = 2 dx^+ dx^- + \sum _{I=1}^{8}\Bigl({dx_I}{dx^I} - \mu^2 x_Ix^I 
(dx^+)^2 \Bigr), \\
\ee
with a constant RR-5 form flux,
\be\la{flux}
F_{+1234}=F_{+5678}=2 \mu,
\ee
where ~$\mu$~ is a scaling parameter of mass dimension one. It is also 
interesting to note that the background in \eq{ppwave}-\eq{flux}, preserves
maximal IIB supersymmetry~\cite{Hull}. What makes 
the new duality tractable, is the fact that string theory in this
background is exactly solvable in light-cone GS formalism~\cite{Metsaev}, 
despite the presence of a nonvanishing RR 
field. Using this 
fact, the authors of~\cite{BMN} gave a proposal to match the string states
of this free world sheet theory, to operators characterizing a certain 
subsector
of the ${\cal{N}} = 4$ super-Yang-Mills theory. The proposal amounts 
to picking a 
~$U(1)_R$~ subgroup of the  ~$SU(4)$~
R-symmetry and constructing trace operators, with a large ~$U(1)_R$~charge 
~${J}$~
and conformal dimension~$\Delta$~, which together scale as~$\sqrt N$~, keeping
the difference~$\Delta - J$~finite in the large~$N$~limit. 
Subsequent
developments included deducing the string interaction vertices from the
three point correlation functions of the these BMN 
operators~\cite{Matthias,Shiraz,Rajesh,Beisert,Santambrogio,
Gross,Gomis,Zhou,Verlinde,Kiem,Berenstein,Huang,Chu1,
Lee,Chu2,Pearson,Gursoy,Plefka}.
Then using light-cone string field theory approach one constructs 
cubic string interaction vertices 
in PP-wave background,
corresponding to splitting or joining of closed strings~\cite{Spradlin1,
Spradlin2,Klebanov,Schwarz1,Pankiewicz,Stefanski,Schwarz2}. 
In this manner, one 
verifies the duality even at the level of interactions.

A generalization of AdS/CFT duality to the one between 
defect CFT and AdS spaces has also been analyzed in the 
literature~\cite{Randall,Ooguri,Bachas,Plee,Ponsot,Erdmenger,Jpark,Sken}. 
Such dualities in general originate from intersecting configurations 
involving a $D3$ brane with other D-branes, when a decoupling 
limit is applied on this brane configuration. 
The conformal structure of the boundary 
theory follows from the fact that the gauge theory living on the 
intersection of the $D3$ and $Dp$-brane goes over in the decoupling limit
to the one living on an $AdS_n \subset AdS_5$. Therefore one also has a 
conformal structure of the $(n-1)$-dimensional  defect CFT, in addition 
to the one coming from $AdS_5$, thus giving a much richer physical structure. 
Considering such an interesting application of $D$-branes to the 
duality between string and gauge theories, it is
important to study $D$-branes as well as their interactions~\cite{Gaberdiel}, 
in  PP-wave backgrounds.
In this context, various brane configurations in PP-wave background 
have been studied.
The world sheet constructions for the free theory on these branes have been 
given in~\cite{ChuHo,Dabholkar,Alok,Alishah,Michishita}. Also,
several supergravity solutions and other aspects of
various brane configurations in PP-wave 
background have been studied in~\cite{Alok,Alishah,Cvetic}.
In addition, emergence of open strings from 
super-Yang-Mills theory has also been shown by 
the authors of~\cite{Maldacena,Vijay}. They construct 
certain determinant and sub-determinant operators , 
which turn out to be the Yang-Mills descriptions of
the $D$-brane states or giant 
gravitons, in  
the PP-wave background. In a recent work~\cite{Taylor,Skenderis}, 
certain symmetry related branes have been found and
are argued to correspond to the giant gravitons. 

In view of the above dualities between open strings and super-Yang-Mills  
at the free theory level, it becomes important to check whether this holds
when the interactions are incorporated. Motivated by the above results,  
in this paper, we construct the 
cubic interaction vertex, corresponding to joining of two open strings to give
a third open string, on a $D7$-brane. To perform this exercise, we use 
the light-cone superstring field theory formalism~\cite{Mand}
developed by Green and Schwarz in flat background~\cite{GreenI,GreenII}. In 
the case of flat space, the transverse ~$SO(8)$~ symmetry 
of the theory was given up in~\cite{GreenIII} and one used 
the spinors of ~$SU(4)$~ for 
certain technical advantages. In the PP-wave background,  
~$SO(8)$~ symmetry is broken 
to ~$SO(4)\times SO(4) $~. $D7$-brane 
boundary conditions break this symmetry further. We therefore use
appropriate decompositions of the space-time coordinates in terms of the
relevant symmetry: $\basis$ to write down vertices and symmetry generators
(for appearance of similar group structure in different context in 
string theory see~\cite{bonelli}). 

This paper is organized as follows. In section-2, we review the world sheet
construction for $Dp$-branes in PP-wave background and give expressions
for all the symmetry generators. In section-3, we discuss the basics of light 
cone string field theory following the work of Green and Schwarz and
construct the cubic interaction vertex involving stringy 
zero modes. In section-4, interaction vertex for all the non-zero open string
modes is derived analogously. In section-5,
we present discussions and conclusions.

\section{Review of worldsheet construction} 

\subsection{$D7$-brane worldsheet theory}

We start by giving the main results of the open string construction 
for the case with both Neumann and Dirichlet  boundary conditions.
We will restrict ourselves to the case of $D7$-branes in this paper, 
although other $D$-branes can be discussed along similar lines. 
The relevant  bosonic degrees of freedom 
in the light-cone gauge Green-Schwarz formalism~\cite{Metsaev,Dabholkar}  are  
$X^I, I=1,\ldots, 8$, which transforms as a vector ${\bf{8_v}}$ of the
transverse $SO(8)$.  In addition, one has space-time 
fermionic degrees of freedom:
$S^1, S^2$ transforming as positive chirality spinors ${\bf 8_s}$ under 
$SO(8)$.  To describe a Dirichlet $p$-brane, we impose Neumann boundary
conditions on $p-1$
coordinates and Dirichlet boundary conditions on the remaining
transverse coordinates:
\bea \la{bbound}
\partial_\sigma X^{r} &=&  0 ,~~~~~   r = 1,...,p-1,\xx 
\partial_\tau X^{r'} &=&  0 , ~~~~~ r' = p,...,8.  
\eea
The class of $D$ branes that we consider here belong to the category  of 
`longitudinal' D-branes in the language of~\cite{Sken}. For such branes  
the light-cone bosonic coordinates : $X^{\pm}$ also satisfy the 
Neumann boundary conditions~\cite{Dabholkar}.
For the fermionic coordinates, the boundary condition is
\bea \la{fbound}
S^1|_{\sigma=0,\pi|\a|} &=& \Omega S^2|_{\sigma=0,\pi|\a|},
\eea
where, as in flat space, $\Omega$ is a real (constant)
matrix $\Pi_k \gamma^k$,
with the product
running over the Dirichlet directions. 
Since, we are interested in $D$-brane 
configurations that preserve sixteen supesymmetries, the choice
of allowed $\Omega$
is constrained further by the following two conditions:
\bea \label{Pi}
[\Omega, \gamma] &=& 0, \xx
\Omega \Pi \Omega \Pi  &=& -1. 
\eea
$\Pi = \gamma^{1234}$ in eqn.\eq{Pi} is the matrix that  defines 
the form of the 
mass term on the worlsheet for these spinors and is responsible for 
breaking the transverse $SO(8)$ symmetry 
down to $SO(4)\times SO(4)$~\cite{BMN}.
The mode expansion of the bosonic coordinates, satisfying the equations of
motion and the boundary conditions in eqns.\eq{bbound}, \eq{fbound} is given by
~\cite{Dabholkar}:
\bea \la{boson}
X^r(\sigma, \tau) &=& x_0^r \cos m\tau + \frac{p_0^r}
{m}\sin m\tau +i\sum_{n
\neq 0}\frac{1}{\omega_n}\a_n^{r} e^{-i\omega_n \tau}
\cos\frac{n\sigma}{|\a|}, \\
X^{r^\prime}(\sigma, \tau) &=& 
\sum_{n \neq 0}\frac{1}{\omega_n}\a_n^{r^\prime} e^{-i\omega_n \tau}
\sin\frac{n\sigma}{|\a|},
\eea
with
\be 
\w_n = {\rm sgn}(n) \sqrt{(n/\a)^2 + m^2},
\ee
and $m=\m p^+$.
For the consistency of the supersymmetry algebra, one chooses the
zero mode of 
$X^{r'}$ to be zero, which corresponds to a $Dp$-brane stuck at
the origin $X^{r'}=0$. 
The mode expansion of the fermions are:
\bea \la{fermions}
S^1 (\sigma,\tau) &=& \cos m\tau S_0
- \sin m\tau \Omega\Pi S_0
+ \sum_{n \neq 0}c_n(\varphi_n^1(\sigma,\tau) \Omega S_n
+i \rho_n \varphi_n^2(\sigma,\tau)\Pi S_n), \xx
\la{s2}
S^2 (\sigma,\tau) &=& \cos m\tau \Omega^T S_0
- \sin m\tau \Pi S_0
+ \sum_{n \neq 0}c_n(\varphi_n^2(\sigma,\tau) S_n
- i \rho_n \varphi_n^1(\sigma,\tau)\Pi \Omega S_n),
\eea
where,
\bea \la{coef}
\phi_n^1 &=  e^{-i(\w_n \tau - \frac {n}{|\a|}\sigma)}, 
\qquad {\phi}_n^2  &=
e^{-i(\w_n\tau + \frac {n}{|\a|}\sigma)},\\
\la{rho}
c_n &=  ( 1+ \rho^2_n)^{-1/2},  \qquad \rho_n &= \frac {\w_n - n/|\a|}{m}.   
\eea
The canonical momenta are:
\be \la{momenta}
P^I~~=~~\frac{1}{2\pi|\a|}\dot{X}^I,~~ I=1,..., 8, \qquad 
{\cal {P}}^{{\cal {I}} a}~~=~~\frac{i}{2\pi|\a|}S^{{\cal {I}} a}, 
~~{\cal {I}}=1, 2,
\ee
and the canonical commutation relations are given in terms of
various modes as:
\bea \la{comm-open}
&&[x_0^r, p_0^s]~=~i\delta^{rs}, \qquad [\a_n^{I},\a_m^{J}]~=~
\frac {1}{2}\omega_n
\delta_{m+n,0} \delta^{IJ}, \\\la{fer-com}
&&\{S_n^{a}, S_m^{b}\}~=~ \frac{1}{4} \delta_{n+m,0}\delta^{ab}.
\eea
For definiteness we now consider the case of a $D7$-brane
extending along
$(+-123456)$ directions. This configuration breaks
the $SO(4)\times SO(4)$
symmetry of the PP-wave background further. Thus we have,
\bea \la{group}
SO(8)&& 
\supset \basis,
\eea
where the $\sul$ and $\sur$ have their origin along 
directions $X^1,...,X^4$ 
and the $\so \times \sop $ come from $X^5,...,X^8$.
Under the embedding in eqn.\eq{group} the spinor decomposes as:
\be \la{spinor}
\bf{8_s} \sim ({\bf 2}, {\bf 1})^{(\half, \half)} \oplus
({\bf {\bar 2}}, {\bf 1})^{(-\half, -\half)} \oplus
({\bf 1}, {\bf 2})^{(\half, -\half)} \oplus
({\bf 1}, {\bf {\bar 2}})^{(-\half, \half)},
\ee
with the superscripts denoting $\so\times\sop$ charges.
Hence, the $\8$ spinors decomposed in terms of the fermionic creation and
annihilation operators are:
\bea \la {fermiond}
\bar{\l}_{\a} &= ~S_{0\a}^{(\half, \half)},\qquad
{\l}_{\a} &= ~S_{0 \a}^{(-\half, -\half)}, \xx
\bar{\l}_{\dot{\a}} &= ~S_{0 \dot{\a}}^{(-\half, \half)},\qquad
{\l}_{\dot{\a}} &= ~S_{0 \dot{\a}}^{(\half, -\half)},
\eea
where $\a$ and $\dot{\a}$ are the doublet indices of $SU(2)_L$ and
$SU(2)_R$ respectively. Although, the decompositions \eq{fermiond} are given 
only for the zero
modes in eqn.\eq{fermions}, similar results hold for the higher modes as well.
The canonical anti-commutation relations in this
basis take the form:
\bea \la{ocsil}
\{\bar{\l}_{\a}, \l^{\b}\} &= \delta_{\a}^{\b}, \qquad
\{\bar{\l}_{\dot{\a}}, \l^{\dot{\b}}\} &= \delta_{\dot{\a}}^{\dot{\b}}.
\eea
The Fock vacuum is,
\be \la{vacuum}
 a_0 |0\rangle = 0,~~~~~ {\l}^{\a} |0 \rangle = 0,
~~~~~ {\l}^{\dot{\a}} |0 \rangle = 0.
\ee
Among the nonzero modes, the positive ones act as annihilation operators and 
the negative ones act as creation operators.

\subsection{Symmetries of the free theory}

We now write the supersymmetry algebra and give 
the expressions for the symmetry
generators for the free theory.
In the light-cone formalism, the generators of the basic
superalgebra can be split into the kinematical generators
$\hat{P}^+$\footnote{We have used a hat over
the momentum generator to avoid mixup with components of momenta defined 
later on.}, 
~~ $P^I,~~J^{+I},~~J^{ij},~~J^{i^\prime j^\prime},~~
Q^{+}_a,~~\bar{Q}^{+}_a$, and the dynamical ones
$H,~~Q^{-1}_{\ad},~~Q^{-2}_{\ad}$.
The dynamical symmetry generators depend on all the non-zero stringy modes 
and also receive corrections from the interactions. 
As has been already mentioned, the left over isometry of the PP-wave
due to the presence of $D7$-branes is $\basis$, and 
the unbroken kinematical symmetriy generators are given as~\cite{Metsaev,
Dabholkar,Taylor}:
\begin{eqnarray} \la{symme}
\hat{P}^+ &=&p^+, \qquad P^r =p_0^r,\qquad J^{+r}= -x_0^r p^+, \\
Q^+&=& \sqrt{2p^+}(1+i\Omega^T)S_0,
\qquad \bar{Q}^+= \sqrt{2p^+}(1-i\Omega^T)S_0, \\
J^{rs}&=& x_0^r p_0^s - x_0^s p_0^r - i S_0 \gamma^{rs} S_0
- i \sum_{n \neq 0} \Bigl\{ \frac{1}{2\omega_n}
(\a_{-n}^r \a_n^s - \a_{-n}^s \a_n^r)
+  S_{-n} \gamma^{rs} S_n \Bigr\},\\
 J^{r^\prime s^\prime} &=& - i S_0 \gamma^{r^\prime s^\prime} S_0
- i \sum_{n \neq 0} \Bigl\{ \frac{1}{2 \omega_n}
(\a_{-n}^{r^\prime} \a_n^{s^\prime} - \a_{-n}^{s^\prime} \a_n^{r^\prime})
+ S_{-n} \gamma^{r^\prime s^\prime} S_n \Bigr\},\\
 2p^+ H &=&  \half(p_{0r}^2 + m^2 x_{0r}^2) - m i S_0 \Omega \Pi S_0
+ \sum_{n \neq 0} \Bigl\{
\half \a_{-n}^I \a_n^I + 2 \omega_n S_{-n} S_n \Bigr\}, \\
\la {q-1}
 \sqrt{2p^+}Q^{-1} &=&  \frac{1}{2\pi \a^\prime p^+}
\int_{0}^{2\pi \a^\prime |p^+|} d\sigma 
\Bigl( \p_- X^I\gamma^I S^1 - m X_I\gamma^I \Pi S^2 \Bigr),\\ 
~~~~~~~~~~  &=&  p_0^r \gamma^r S_0 + m x_0^r \gamma^r \Omega\Pi S_0
- \sum_{n \neq 0} \Bigl\{ c_n \a_{-n}^I \Omega \gamma^I S_{n}
- \frac{i m}{2c_n \omega_n}\a_{-n}^I \gamma^I \Pi S_{n} \Bigr\}, \\
\la {q-2}
\sqrt{2p^+}Q^{-2} &=& \frac{1}{2\pi \a^\prime p^+}
\int_{0}^{2\pi \a^\prime |p^+|} d\sigma \Bigl( \p_+ X^I\gamma^I
S^2 + m X_I\gamma^I \Pi S^1 \Bigr),\\
~~~~~~~~~~  &=& p_0^r \gamma^r \Omega^T S_0 + m x_0^r \gamma^r \Pi S_0
+ \sum_{n \neq 0} \Bigl\{ c_n \a_{-n}^I \gamma^I S_{n}
- \frac{i m}{2c_n \omega_n}\a_{-n}^I \Omega^T \gamma^I \Pi S_{n}
\Bigr\},
\end{eqnarray}
Since, $Q^+$ and $\bar{Q}^+$ as well as $Q^{-1}$ and
$Q^{-2}$ are related as~\cite{Dabholkar},
\begin{equation}
Q^+ + Q^- + i \Omega (Q^+ - Q^-) =0, \qquad
Q^{-1}- \Omega Q^{-2} =0,
\end{equation}
one generally considers the particular combinations given below,
which preserve
the $D7$-brane 
supersymmetries that we are interested in:
\begin{eqnarray} \la{q's}
q^+ &=& \half \Bigl(Q^+ + Q^- - i \Omega (Q^+ - Q^-) \Bigr)
= 2\sqrt{2p^+} S_0, \\
\la{qminus}
q^- &=& Q^{-1} + \Omega Q^{-2} = 2 Q^{-1}.
\end{eqnarray}
The non-vanishing (anti-)commutation relations are~\cite{Dabholkar}:
\begin{eqnarray} \la{open-jq}
&&[\hat{P}^-, P^I] = \mu^2 J^{+I},~~~
\qquad [P^I, J^{+J}] = - \delta^{IJ} \hat{P}^+,\\  
&& [\hat{P}^-, J^{+I}] = P^I,~~~ 
\qquad [J^{rs}, q^{\pm}]=\frac{i}{2} \gamma^{rs} q^{\pm},\\
&& [J^{r^\prime s^\prime}, q^{\pm}]=
\frac{i}{2} \gamma^{r^\prime s^\prime}q^{\pm},~~~ 
 [J^{+r}, q^-]= \frac{i}{2} \Omega^T \gamma^r q^+, \\
&& [P^r, q^-]=-i \frac{\mu}{2 p^+} \gamma^r \Pi q^+,
\qquad [H, q^+]= i \frac{\mu}{2p^+} \Omega \Pi q^+, \\
&& \{q^+_{\a}, q^+_{\b}\}=\delta_{ab} 2\hat{P}^+, \\
&& \{q^+_{\a}, q^-_{\ad}\}= (\Omega \gamma^r)_{a\ad}  P^r - \frac{\mu}{p^+}
(\Pi \gamma^r)_{a\ad} J^{+r}, \\
 \la{dynq}
&& \{q^-_{\ad}, q^-_{\bd}\}= \delta_{\ad\bd} 2(H) + \frac{\mu}{2p^+}
\Bigl( (\gamma^{rs}_I \Pi\Omega)_{\ad\bd} J^{rs}_I +
(\gamma^{r^\prime s^\prime}_I \Pi\Omega)_{\ad\bd}
J^{r^\prime s^\prime}_I
\Bigr) \xx
&& \qquad \qquad \;\; - \frac{\mu}{2p^+}
\Bigl( (\gamma^{rs}_{II} \Pi\Omega)_{\ad\bd} J^{rs}_{II} +
(\gamma^{r^\prime s^\prime}_{II} \Pi\Omega)_{\ad\bd}
J^{r^\prime s^\prime}_{II}
\Bigr),
\end{eqnarray}
where $J^{rs}_I \in SO(m), J^{r^\prime s^\prime}_I \in SO(4-m),
J^{rs}_{II} \in SO(n)$, and $J^{r^\prime s^\prime}_{II} \in SO(4-n)$.
In section-3, we will rewrite some of these generators and the 
commutation algebra using the symmetry structure in eqn.\eq{group}. 

\section{Zero mode vertex in GS light-cone SFT}

In this section we first discuss some known results of the GS light-cone string
field theory formalism and then apply them to write down the zero mode 
interaction vertex.
As we shall see, most of the conceptual and technical aspects of the problem
are already addressed in the zero mode analysis.
We start by discussing the cubic open-string vertex 
corresponding to a process, 
where two open strings on a $D7$-brane
join at their ends at $\tau = 0$ to give a third open string. The construction
follows the earlier work of Green 
and Schwarz~\cite{GreenI,GreenII,GreenIII} for
flat space. It involves writing down the 
dynamical generators in terms of the string fields, which can create or destroy
complete strings.  The string fields can be expanded in a number basis
representation as follows,
\begin{equation} \la{sfield}
\Phi[p(\sigma)]
= \sum_{ \{ n_k \} } \varphi_{\{ n_k \}} \prod_{k=-\infty}^\infty
\psi_{n_k}(p_k),
\end{equation}
where $p_k$ is the $k$-th Fourier mode of $p(\sigma)$ and
the sum is over all possible sets of harmonic oscillator
occupation numbers $\{n_k\}$ with
$\psi_n(p)$ being the harmonic oscillator wavefunction for
occupation number $n$. Here, $\varphi_{\{n_k\}}$ is an operator  that
creates or destroys a string in the state $|\{n_k\}\rangle$ at time $\tau = 0$.
The full Hamiltonian has the form
\be \la{hamil}
H = H_2 + \kappa H_3 + ...,
\ee
with similar expansions for $Q^-$ and $\overline{Q}^-$,
where $\kappa$ is the  coupling constant. The interaction Hamiltonian as well 
as other dynamical generators can be expressed in terms of the 
string fields and 
hence, in a number basis representation.
The coordinates of the three strings are parametrized as follows,
\bea \la{sigmas}
\s_{(1)}=& \s & {\rm for} ~~~0 \leq \s \leq \pi \a_{(1)},\cr
\s_{(2)}=& \s - \pi \a_{(1)} & {\rm for}~~~ 
\pi \a_{(1)} \leq \s \leq \pi (\a_{(1)} + \a_{(2)}),\cr
\s_{(3)}=& \s- \pi(\a_{(1)} + \a_{(2)}) & {\rm for}~~~
0 \leq \s \leq \pi (\a_{(1)} + \a_{(2)}),
\eea
where $\a_{r} = 2\a' p^+$, and $r = 1,2,3$ denote the string indices . We also
have $\a_{(1)} + \a_{(2)} + \a_{(3)} = 0$, and 
one takes $\alpha_{(1)}$,
$\alpha_{(2)}$ positive as the corresponding strings are annihilated by the 
vertex operator. With this labelling, the coordinates of the strings mean,
\be 
X_r(\sigma) = x^{(r)}(\sigma _r) \Theta _r ,
\ee
where
\be \la{aaa}
\Theta_{(1)} = \theta(\pi \alpha_{(1)} - \sigma), \qquad \Theta_{(2)}
= \theta(
\sigma - \pi \alpha_{(1)}), \qquad \Theta_{(3)} = 1.
\ee
The rest of the coordinates can be written in a similar form. 

Before proceeding, it is important to decompose 
the $SO(8)$ invariant dynamical 
supercharges in the $SU(2)$ doublet notation according 
to the embedding \eq{group}.
Under this embedding, the vectors decompose as,
\bea \la{vector}
\bf{8_v} &\sim& ({\bf 2}, {\bf 2})^{(0, 0)} \oplus
({\bf 1})^{(+1,0)}\oplus {\bf (\bar 1})^{(-1, 0)}\oplus 
({\bf 1})^{(0,+1)}\oplus {\bf (\bar 1})^{(0, -1)}
\eea
and the fermion decomposition is given in eqn.\eq{spinor},
where the superscripts indicate the $\so\times\sop$ charges corresponding
to rotations in 56 and 78 directions.
To proceed further, we need various expressions written in terms of the 
representations of the group $\basis$.
Therefore, the relevant string degrees of 
freedom for us (e.g., eqn.\eq{s2}) are,
\bea \la{S0matrix}
S_0 &=& \pmatrix{\lambda^1_{\ad}  \cr
                 \lambda^1_{\a} \cr
                        \lambda^2_{\ad} \cr
                        \lambda^2_{\a}  },
\eea
with ~$\lambda_{\a} = \lambda^1_{\a}-i\lambda^2_{\a}$, 
$\lambda_{\ad} = \lambda^1_{\ad}-i\lambda^2_{\ad}$~  and ~$\lambda_{\a}
,\lambda_{\ad} $ ~as given in eqn.\eq{fermiond}.
Similarly, in view of eqn.\eq{vector}, the bosonic coordinates $X^I$ and and 
their conjugate momenta $P^I$ with $I= 1,...8$, can be decomposed in terms of:
$X_{\ad\a}~,X^{\pm},\tilde{X}^{\pm}$ and 
$P_{\ad\a}~,P^{\pm},\tilde{P}^{\pm}$,
where $ X_{\ad\a}$  and $P_{\ad\a}$ are complex combinations of the coordinates
$x^{1,...4}$ and $p^{1,...4}$ as given below:
\be \la{matrix}
X_{\ad \a}~=~ \pmatrix{x^4 + ix^2 & -x^3 + i x^1 \cr
                                  x^3 + i x^1 & x^4 - i x^2 },\qquad
P_{\ad \a}~=~ \pmatrix{p^4 + ip^2 & -p^3 + i p^1 \cr
                                  p^3 + i p^1 & p^4 - i p^2 },
\ee
with $\a$ and $\ad$ being the $\sul$, $\sur$ indices respectively. 
Further, we have the definitions,
$X_{\a\ad} ~=~X_{\ad \a}^T~,~\bar X_{\a\ad}~=~X_{\ad\a}^{\dagger}$, and similar
ones for $P_{\a \ad}$ and  $\bar P_{\a\ad}$.
Also, $X^{\pm},\tilde{X}^{\pm}$ and $P^{\pm},\tilde{P}^{\pm}$ are 
defined as:
\bea
&&X^{\pm}~=~ x^6~\pm~ix^5,~ \qquad \tilde{X}^{\pm}~=~
\tilde{x}^8~\pm~i\tilde{x}^7,\\
\la{pptilde}
&&P^{\pm}~=~ p^6~\pm~ip^5,~ \qquad \tilde{P}^{\pm}~=~
\tilde{p}^8~\pm~i\tilde{p}^7.
\eea
Above combinations of coordinates and momenta have in fact 
been obtained by using
a specific representation of 8-d Dirac $\gamma$ matrices:
\bea
&&\gamma^1 ~=~ \ep\otimes  I\otimes  \t_1,
\qquad~\gamma^2 ~=~ \ep\otimes  I\otimes \tau_3,
\qquad~\gamma^3 ~=~  I\otimes  \t_3\otimes  \ep,\\
&&\gamma^4 ~=~ I\otimes  I\otimes I,
\qquad \;\;\gamma^5 ~=~ \ep\otimes \ep\otimes \ep,
\qquad \;\;\;\gamma^6 ~=~  I\otimes  \t_1\otimes  \ep,\\
&&\gamma^7 ~=~ \t_1\otimes  \ep\otimes  I,
\qquad \;\;\gamma^8 ~=~ \tau_3\otimes \ep \otimes I,
\eea
where $\ep = i\t_2$. These $\gamma$'s which decompose the quantities
$X^I\g^I$ and $P^I\g^I$ as in 
eqn.\eq{matrix}-\eq{pptilde} are identical to the ones in~\cite{GS}, 
except for 
minor relabelling
suitable to match with results in~\cite{Dabholkar}. In this representation of
$\gamma$ matrices $\O$ and $\Pi$ are given by,
\be \la{omegapi}
\O~=~ \pmatrix{ 0 & I_4 \cr
                             -I_4 & 0 }, \qquad
\Pi~=~ \pmatrix{ I_2\otimes\t_3 & 0\cr
                                0 & I_2\otimes\t_3 }. 
\ee
Using the above results, supersymmetry generators are written in terms of 
$\lambda_{\a},\lad,\lba,\lbad$. The 
kinematical supersymmetry generators for example have the form:
\bea \la{Qplus}
&&Q^+~=~ \pmatrix{\bl_{\ad}  \cr
                 \bl_{\a} \cr
                        -i\bl_{\ad} \cr
                        -i\bl_{\a}  },
\qquad
\bar{Q}^+~=~ \pmatrix{\l_{\ad}  \cr
                 \l_{\a} \cr
                        i\l_{\ad} \cr
                        i\l_{\a}  },
\eea
with invariant combinations in eqn.\eq{q's} given by eqn.\eq{S0matrix}.
In eqn.\eq{Qplus}, we have absorbed a factor of $\sqrt{2p^+}$ in the 
redefinition of $\lambda$'s, in order to make connection with results 
of~\cite{GreenIII}.

Now, the procedure for constructing the vertex consists of two steps. The first
step amounts to obtaining the kinematical part of the three string vertex that
would occur, if the entire
interaction Hamiltonian were given only by the kinematical factors and no other
terms. These are the same as in flat space and arise from the commutation 
relations of dynamical generators $H$ with the kinematical generators 
$P^I,J^{+I},q^+$. This gives the exponential part of the vertex. We postpone
more detailed discussion of this part to section-4. Here, our main focus 
would be on the second step, which
involves the calculation of prefactor, obtained
from the algebra of dynamical generators. 

At the level of zero modes, the kinematical
part of the vertex takes a simple form and can be obtained by
imposing the constraints 
coming from the PP-wave superalgebra of kinematical generators.
Hence, the bosonic and fermionic
coordinate continuity and local momentum conservation equations take the form:
\bea 
\la{kin2} \Bigl(\alpha_{(1)}
\hat{x}_{(1)} + \alpha_{(2)} \hat{x}_{(2)} + \alpha_{(3)} \hat{x}_{(3)}
\Bigr)\V &=& 0,\\
\la{kin1}
\Bigl(\hat{p}_{(1)} + \hat{p}_{(2)} + \hat{p}_{(3)}\Bigr)\V &=& 0, \\
\la{kin3}
\Bigl(\alpha_{(1)}\hat{\l}_{(1)}+\alpha_{(2)}\hat{\l}_{(2)}+ \alpha_{(3)}
\hat{\l}_{(3)}\Bigr)_{\a,\ad}\V &=& 0,\\
\la{kin4}
\Bigl(\hat{\bl}_{(1)} + \hat{\bl}_{(2)} + \hat{\bl}_{(3)}
\Bigr)_{\a,\ad}\V &=& 0.
\eea
Note that the indices (which we have suppressed)
on coordinates and momenta run only over the 
Neumann directions (having zero modes), 
and hence the constraints on the vertex are only in these directions.  

Hence, the kinematical part of the zero mode vertex looks as follows:
\bea \la{zeroV}
|V_{kin}\ra &=& |E^0_b\ra |E^0_f\ra, 
\eea
where the bosonic part,
\bea \la{zerobosV}
|E^0_b\ra &\sim& \exp \Bigl[~\frac{1}{2} \sum_{r,s = 1}^3 a_{(r)}^\dagger
M^{rs}a_{(s)}^\dagger~\Bigr]|0\ra,
\eea
and the fermionic part,
\bea
\la{zeroferV}
|E^0_f\ra &\sim& {1 \over 4!}\eab\eabd
\hat{\bl}^{\a}\hat{\bl}^{\b}\hat{\bl}^{\ad}\hat{\bl}^{\bd}|0\rangle, 
\eea
with $\hat{\bl} = \hat{\bl}_{(1)}+\hat{\bl}_{(2)}+\hat{\bl}_{(3)}$ and
$a^{\dagger}=1/\sqrt{\a\m}~\hat{p}_0 + i/2\sqrt{\a\m}~\hat{x}_0$.
The matrix $M^{rs}$ is as defined in~\cite{Spradlin1}, 
and is being given in the appendix.
Having determined the kinematical part of the vertex $\V$, we now
calculate the prefactor for the zero modes of the open 
strings ending on the $D7$-brane.

The calculation of prefactor involves making an ansatz, for the functions
multiplying the kinematical part of the vertex and the dynamical supercharges. 
The algebra of dynamical 
generators, then provides certain consistency conditions. These conditions
determine the unknown functions uniquely.
We begin by writing down the independent components of the dynamical 
supersymmetry generators, which can be seen by examining either 
$Q^{-1}$ or $Q^{-2}$. 
Using the
decompositions \eq{spinor} and \eq{vector}, the dynamical supercharge given
in eqn.\eq{q-1} (or equivalently eqn.\eq{qminus}) can be written in the 
following form:
\bea \la{a}
q^{-}_{\a} &=& \frac {1}{\a}~\Bigl[~P_{\a\ad} \l_{\ad} +
P^+(\ep \l)_{\a} ~\Bigr]
~-~i\mu~\Bigl[~X_{\a\ad} \l_{\ad} -  X^+(\ep \l)_{\a}~\Bigr],\\
\bq ^{-}_{\a} &=& \frac {1}{\a}~\Bigl[~ \bp _{\a\ad} \bl_{\ad} + 
P^-(\ep \bl)_{\a} ~\Bigr]~+~i\mu~\Bigl[~ \bx_{\a\ad} \bl_{\ad} - 
 X^-(\ep \bl)_{\a}~\Bigr],\\
q^{-}_{\ad} &=& \frac {1}{\a}~\Bigl[~P_{\ad \a} \l_{\a} +
P^-(\ep \l)_{\ad} ~\Bigr]
~+~i\mu~\Bigl[~X_{\ad \a} \l_{\a} -  X^-(\ep \l)_{\ad}~\Bigr],\\
\bq^{-}_{\ad} &=& \frac {1}{\a}~\Bigl[~\bp_{\ad \a} \bl_{\a} +
P^+(\ep \bl)_{\ad}
~\Bigr]
~-~i\mu~\Bigl[~\bx_{\ad \a} \bl_{\a} -  X^+(\ep \bl)_{\ad}~\Bigr].
\eea
Now, we use the above form of generators to determine the interaction vertex
$|H\ra,|\qa\ra$ etc., in a number basis defined in terms of the string fields
as~\cite{GreenIII,Spradlin1}:
\be
\langle 3 | H_3 | 1 \rangle |2 \rangle
=  \langle 1 | \langle 2 | \langle 3|H\rangle.
\ee
The kinematical part of the vertex has been already given in 
eqn.\eq{zeroV}-\eq{zeroferV}. The prefactors multiplying the kinematical
vertex are given in terms of the unknown functions (to be determined below) 
$f_{\a\ad},\bif_{\a\ad},f^{\pm},f_{\a},f_{\ad},\bar{f}_{\a},\bar{f}_{\a}$ as: 
\bea \la{h}
|~H~\ra &=& \Bigl[~\IP_{\a\ad}~f^{\a\ad}~+~\bip_{\a\ad }
~\bif^{\a\ad}~+~\IP^+~f^-~+~\IP^-~f^+~\Bigr]~\V, \\
\la{dynqminus}
|~q^-_{\a}~\ra &=& f_{\a}~\V,\\
|~q^-_{\ad}~\ra &=& f_{\ad}~\V, \\
|~\bq^-_{\a}~\ra &=& \bif_{\a}~\V, \\
\la{dynqbarminus}
|~\bq^-_{\ad}~\ra &=& \bif_{\ad}~\V,
\eea
where,
\bea
\pada&=&~\a_{(1)}~P_{\ad\a(2)}~-~\a_{(2)}~P_{\ad\a(1)},\\
\Lad&=&~\a_{(1)}~\lbad_{(2)}~-~\a_{(2)}~\lbad_{(1)},
\eea
with similarly looking expressions for $\pbaad,P^{\pm},\Lambda_{\a},
\Lambda_{\ad}$ and $\La$. The unknown quantities $f_{\a},\bar{f}_{\a}$ etc.,
multiplying the vertex 
are assumed to be functions of 
only $\bL$, as ~$\L \V = 0$. In other words, they 
are functions of the bosonic and
fermionic conjugate momenta only. Above ansatz for the prefactors are slightly 
different from the ones in~\cite{GreenII,Spradlin1} for the closed strings and 
are along the lines of 
results for open string in flat space~\cite{GreenIII}. 
The form of functions $f$ 
can now be
determined by requiring that the PP-wave superalgebra is satisfied to 
$O(\k)$. For the generators $\qa$ transforming nontrivially under the 
group $\sul$ in eqn.\eq{dynq}, this takes the form:
\bea \la{pp1h} 
\sum_{r=1}^3 ~q^-_{\a(r)}~
|~\overline{q}^-_{\b}~\ra 
~+~ \sum_{r=1}^3 ~\overline{q}^-_{\b(r)}~
| ~q^-_{\a}~\ra 
~&=&~ 2~ \delta_{\a\b}~ |~H~\ra,\\
\la{pp1qm}
\sum_{r=1}^3~ q^-_{\a(r)}~ |~ q^-_{\b}~\ra
~+~ (\a\leftrightarrow \b) ~&=& ~0,\\
\la{pp1qb}
\sum_{r=1}^3~ \overline{q}^-_{\a(r)}~ |~ \overline{q}^-_{\b}~\ra
~+~ (\a\leftrightarrow \b) ~&=&~ 0.
\eea
One has identical relations for the $q_{\ad}'s$ as well. The remaining 
ones between $\qa$ and $\qad's$ are:
\bea \la{pp2}
\sum_{r=1}^3~ q^-_{\a(r)}~ |~ q^-_{\ad}~\ra
~+~ (\a\leftrightarrow \ad) ~&=& ~0,\\
\la{aabard}
\sum_{r=1}^3~ q^-_{\a(r)}~ |~ \bq^-_{\ad}~\ra
~+~ (\a\leftrightarrow \ad) ~&=& ~0,\\
\la{abarad}
\sum_{r=1}^3~ \bq^-_{\a(r)}~ |~ q^-_{\ad}~\ra
~+~ (\a\leftrightarrow \ad) ~&=& ~0,\\
\la{abarabard}
\sum_{r=1}^3~ \overline{q}^-_{\a(r)}~ |~ \overline{q}^-_{\ad}~\ra
~+~ (\a\leftrightarrow \ad) ~&=&~ 0.
\eea
Also, the canonical commutation relations in our basis take the form:
\bea \la{bc}
&&\Bigl[~X_{\a\ad}~,~\bar{P}_{\b\bd}~\Bigr]~=~
i~\delta_{\a\b}~\delta_{\ad \bd}~\delta_{rs}~,
~~~~\qquad \Bigl[~\bx_{\a\ad}~,~P_{\b\bd}~\Bigr]~=~
i~\delta_{\a\b}~\delta_{\ad \bd}~\delta_{rs}~,\\
&&\Bigl[~X^{\pm}~,~P^{\mp}~\Bigr]~~~=~ i~\delta_{rs}~,~~~~~~~~~~~~~
\qquad \qquad \Bigl[~\xt^{\pm}~,~\pt^{\mp}~\Bigr]~=~ i~\delta_{rs}~,\\
\la{fa}
&&\Bigl\{~\bl_{\a(r)}~,~\lambda^{\b}_{(s)}~\Bigr\}~~ =~
\a_r~\delta_{\a}^{\b}~\delta_{rs}~,~~~~
\qquad \Bigl\{~\bl_{\ad(r)}~,~\lambda^{\bd}_{(s)}~\Bigr\}~
=~\a_r~\delta_{\ad}^{\bd}~\delta_{rs},
\eea
with r and s being string indices. Further, one gets the following useful 
set of relations involving the supercharges $\qa$:
\bea \la{qacomm}
\sum_{r=1}^3~ \Bigl[~\qa~,~\bar{\IP}_{\b\bd}~\Bigr]~&
=&~\m\L_{\ad}\dab\dabd,\\
\sum_{r=1}^3~ \Bigl[~\qa~,~\IP^- ~\Bigr]~&=&~ -\m~(\ep \Lambda)_{\a},\\
\sum_{r=1}^3~ \Bigl\{~\qa~,\Lb\Bigr\}~&=&~
\Bigl(~\IP^+~-~i\mu\a~\IR^+~\Bigr)\eab ,\\
\sum_{r=1}^3~ \Bigl\{~\qa~,\Lbd\Bigr\}~&=&~
\Bigl(~\IP_{\a \ad}~+~i\m\a~\IR_{\a\ad}~\Bigr)\dabd,
\eea
and the following algebra for the complex conjugates $\qab$:
\bea
\sum_{r=1}^3~ \Bigl[~\qab~,~\IP_{\b\bd}~\Bigr]~&=&~-\m\Lbd\dab,\\
\sum_{r=1}^3~\Bigl[~\qab~,~\IP^+~\Bigr]~&=&~\mu(\ep \bL)_{\a},\\
\sum_{r=1}^3~ \Bigl\{~\qab~,~\L_{\bd}\Bigr\}&=&\Bigl(\bip_{\a\ad}~-
~i \m\a~\bir_{\a\ad}~\Bigr)\dabd,\\
\sum_{r=1}^3~ \Bigl\{~\qab~,~\L_{\b}~\Bigr\}~&=&~
\Bigl(~\bin~+~i\mu\a~\IR^{-}~\Bigr)\eab,
\eea
where $\IR_{\a\ad}~=~\frac{1}{\a_{3}}~(~X_{(2)\a\ad}~-~X_{(1)\a\ad})$, with 
$\bir_{\a\ad},\IR^{\pm}$ defined analogously. 
Similarly, for the $\qad's$, we have:
\bea \la{qadcomm}
\sum_{r=1}^3~ \Bigl[~\qad~,~\bip_{\bd \b}~\Bigr]~&=&~-\m\L_{\b}\dabd,\\
\sum_{r=1}^3~\Bigl[~\qad~,~\IP^+~\Bigr]~&=&~ \m~(\ep \Lambda)_{\ad},\\
\sum_{r=1}^3~ \Bigl\{~\qad~,\Lbd\Bigr\}~&=&~ \Bigl(~\IP^-~ 
+~i\mu\a~\IR^-~\Bigr)\eabd ,\\
\sum_{r=1}^3~ \Bigl\{~\qad~,\Lb\Bigr\}~&=&~\Bigl(~\IP_{\ad \a}~
-~i\m\a~\IR_{\ad \a}~\Bigr)\dab.
\eea
The relations involving the complex conjugates $\qabd$ are given as: 
\bea
\sum_{r=1}^3~ \Bigl[~\qabd~,~\IP_{\bd \b}~\Bigr]~&=&~\m\Lb\dabd,\\
\sum_{r=1}^3~ \Bigl[~\qabd~,~\IP^-~\Bigr]~&=&~-\mu(\ep \bL)_{\ad},\\
\sum_{r=1}^3~ \Bigl\{~\qabd~,~\L_{\bd}~\Bigr\}~&=&~
\Bigl(~\bim~-~i\mu\a~\IR^{+}~\Bigr)\eabd, \\
\la{qdlast}
\sum_{r=1}^3~ \Bigl\{~\qabd~,~\L_{\b}~\Bigr\}~&=&~\Bigl(~\bip_{\ad \a}~+~
i \m\a~\bir_{\ad \a}~\Bigr)\dab.
\eea
Using the conservations laws in eqn.\eq{kin2}-\eq{kin4} the state $\V$ 
can be shown to satisfy:
\bea \la{Ppv}
\sum_{r=1}^3~\qa~\V~&=&~0,\\
\sum_{r=1}^3~\qad~\V~&=&~0,\\
\sum_{r=1}^3~\qab~\V~&=&~-\frac{1}{\a}~\Bigl
[~\bip_{\a\ad}\Lad~+~\bin(\ep \bL)_{\a}~\Bigr]
~\V,\\ \la{pplast}
\sum_{r=1}^3~\qabd~\V~&=&~-\frac{1}{\a}~\Bigl
[~\bip_{\ad \a}\La~+~\bim(\ep \bL)_{\ad}~\Bigr]~\V.
\eea
Now, substituting our ansatz for the prefactors of the dynamical generators
given in eqn.\eq{h}, in the PP-wave superalgebra in eqn.\eq{pp1qm},
\eq{pp1qb} and using the above relations, we
end up with the following equations for the unknown functions 
$f_{\a},\bar{f}_{\a}$:
\bea \la{ffirst}
\frac{\p f_{\b}}{\p\Lc} \eac~+~\frac{\p f_{\a}}{\p\Lc} \ebc~&=& ~0,\\ 
\frac{\p f_{\b}}{\p\Lcd} \paad\dacd~+~\frac{\p f_{\a}}{\p\Lcd}~
\pbbd\dbcd ~&=&~0,\\
\bar{f_{\b}}\Lc \eac~+~\bar{f_{\a}}\Lc \ebc~&=&~0,\\ 
\bar{f_{\b}}\pbaad\Lad ~+~\bar{f_{\a}}\pbbbd\Lbd~&=& ~0.
\eea
These relations are obtained by comparing the coefficients of 
$\paad,\pbaad$~and~ $P^{\pm}$ on both sides of eqns.\eq{pp1qm} and \eq{pp1qb}.
Similar equations are satisfied by their dotted counterparts $f_{\ad}$ and
$\bar{f}_{\ad}$.
There are additional conditions coming from the
set of algebra in eqns.\eq{pp2}-\eq{abarabard}. The relation in
eqn.\eq{pp2} leads to the following equations involving $f_{\a}$ and $f_{\bd}$:
\bea
\frac{\p f_{\bd}}{\p \Lc}\eac &=&~0,\\
\frac{\p f_{\a}}{\p \Lcd}\ebcd &=&~0,\\
\frac{\p f_{\bd}}{\p \Lcd}\paad\dacd~+~\frac{\p f_{\a}}{\p \Lc}\pbdb \dbc ~&=&~0.
\eea
Similarly, the relation in eqn.\eq{aabard} gives the following set of equations
for the unknowns $f_{\a}$ and $\bar{f}_{\bd}$:
\bea
\frac{\p\bar{f_{\bd}}}{\p\Lc}\eac~+~\frac{1}{\a}f_{\a}~(\ep\bL)_{\bd} ~&=&~0,\\
\frac{\p\bar{f_{\bd}}}{\p\Lcd}\paad\dacd~+~
\frac{1}{\a}f_{\a}~\pbbdb\Lb~&=&~0.
\eea
The set of algebra in eqns.\eq{abarad}
result in the following consistency conditions for $f_{\bd}, \bar{f}_{\a}$:
\bea
\frac{\p\bar{f_{\a}}}{\p\Lcd}\ebcd ~+~ \frac{1}{\a}f_{\bd}~
(\ep\bL)_{\a}~&=&~0,\\
\la{flast}
\frac{\p\bar{f_{\a}}}{\p\Lc}\pbdb\dbc ~+~ \frac{1}{\a}f_{\bd}~
\pbaad\Lad~&=&0.
\eea
and finally the relations in eqn.\eq{abarabard} lead 
to the following conditions 
on $\bar{f}_{\a},\bar{f}_{\bd}$:
\bea
\frac{1}{\a}\bar{f_{\bd}}(\ep\bL)_{\a}~&=&~0,\\
\frac{1}{\a}\bar{f_{\a}}(\ep\bL)_{\bd}~&=&~0,\\
\bar{f_{\bd}}\pbaad\Lad ~+~\bar{f_{\a}}\pbbdb\Lb~&=&0,
\eea
It is worth noting that some care has to be taken while solving the
above consistency conditions, as they 
involve grassmann functions. We find that the 
set of equations \eq{ffirst}-\eq{flast}, are solved by the functions:
\bea \la{sol1}
f_{\a}~&=&~ -2~\La,\\
f_{\ad}~&=&~2~\Lad,\\
\bar{f_{\a}} ~&=&~\frac{1}{\a}\eab\Lb\ep_{\cd\dd}\Lcd\bL_{\dd},\\
\la{sol4}
\bar{f_{\ad}} ~&=&~-\frac{1}{\a}\eabd\Lbd\ep_{\ga\d}\Lc\bL_{\d}.
\eea
The solutions in eqns.\eq{sol1}-\eq{sol4} complete the 
determination of prefactors 
appearing in the ansatz \eq{dynqminus}-\eq{dynqbarminus} for the dynamical
supercharges. We now go
on to determine the prefactors of the interaction hamiltonian and hence, the
cubic open-string vertex operator. Substituting the ansatz given in eqn.\eq{h}
for the interaction hamiltonian in eqn.\eq{pp1h}, we end up with the following
relations:
\bea \la{h1}
\frac{\p \bar{f_{\b}}}{\p \Lc}\eac &=&~2~\dab f^-,\\ 
\frac{1}{\a}f_{\a}\ebc\Lc &=&~2~\dab f^+,\\ \la{h4}
\frac{\p\bar{f_{\b}}}{\p\Lcd}\paad\dacd ~+~ \frac{1}{\a}f_{\a}~\pbbbd\Lbd~&=&~
2\dab\Bigl(~\IP_{\ga\dot{\gamma}}~f^{\ga\dot{\ga}}~+~\bip_{\ga\dot{\ga}}
{\bar f}^{\ga\dot{\ga}}~\Bigr).
\eea
Now, using eqns.\eq{sol1}-\eq{sol4} in eqns.\eq{h1}-\eq{h4}, the 
solutions for the unknown functions in the ansatz 
for the interaction hamiltonian 
\eq{h} are given as $f_{\a},\bar{f}_{\a}$:
\bea \la{hsol1}
f^-~&=&~\frac{1}{2\a}\eabd\Lad\Lbd,\\
f^+~&=&~-\frac{1}{2\a}\eab\La\Lb,\\
f_{\a \ad}~&=&~-\frac{1}{\a}\eac\ep_{\ad\dd}\Lc\bL_{\dd},\\
\la{hsol4}
{\bar{f}_{\a\ad}}~&=&~-\frac{1}{\a}\La\Lad.
\eea
The dotted counterpart to equation (\ref{pp1h}) leads to slightly 
different conditions than in eqns. (\ref{h1})-(\ref{h4}). However, the 
final answer for the prefactor of the interaction hamiltonian turns out to be 
identical to the ones in eqns. (\ref{hsol1})-(\ref{hsol4}).

Hence, the solutions given in 
eqns.\eq{sol1}-\eq{sol4} and \eq{hsol1}-\eq{hsol4} 
determine the complete cubic 
interaction vertex and the supercharges 
for an open string on a $D7$-brane at the level of zero modes when substituted 
in eqns.\eq{h}-\eq{dynqbarminus}.
\vskip 0.3cm
\section{Superstring vertex}
In this section we generalize our results of the previous section to the
full string theory by following the approach of~\cite{GreenIII,Spradlin1} 
for both closed and open strings earlier. 
Before proceeding, we note, that at $\t=0$ the bosonic mode 
expansions for Neumann directions given in 
eqns.\eq{boson} can be written in the form:
\bea \la{xt0}
X^r(\sigma) &=& x_0^r +i\sum_{n=1}^{\infty}\Bigl[\frac{1}{\omega_n}
(\a_n^{r} - \a_{-n}^{r})\cos\frac{n\sigma}{|\a_|}\Bigr], 
\eea
and their conjugate momenta are given as :
\bea \la{pt0}
P^r(\sigma) &=& \frac{1}{\pi|\a|}\Bigl[p_0^r +\sum_{n=1}^{\infty}
(\a_n^{r} + \a_{-n}^{r})\cos\frac{n\sigma}{|\a_|}\Bigr].
\eea
For Dirichlet directions we have:
\bea \la{xtd}
X^{r'}(\sigma) &=& \sum_{n=1}^{\infty}\Bigl[\frac{1}{\omega_n}
(\a_n^{r'} + \a_{-n}^{r'})\sin\frac{n\sigma}{|\a_|}\Bigr], 
\eea
with their conjugate momenta being:
\bea \la{ptd}
P^{r'}(\sigma) &=& \frac{1}{\pi|\a|}\Bigl[-i\sum_{n=1}^{\infty}
(\a_n^{r'} - \a_{-n}^{r'})\sin\frac{n\sigma}{|\a_|}\Bigr].
\eea

To write down the mode expansions of fermions $S^1$ and $S^2$, 
one can use a decomposition similar to the one in previous section 
for zero modes, i.e. eqn. (\ref{S0matrix}). One also notes using 
eqns. (\ref{fermions}) that they satisfy $S^1(\s) = \Omega S^2(- \s)$.
We therefore use the following decompositions for the fermion fields and
their modes in eqn. (\ref{fermions}) :
\bea \la{Smat}
S^2(\s) &=~~ \pmatrix{\l^1_{\ad}(\s)  \cr
                 \l^1_{\a}(\s) \cr
                       \l^2_{\ad}(\s) \cr
                        \l^2_{\a}(\s)  },\qquad
S_n &=~~ \pmatrix{\lambda^1_{n\ad}  \cr
                 \lambda^1_{n\a} \cr
                        \lambda^2_{n\ad} \cr
                        \lambda^2_{n\a}  }.
\eea
The complex combinations $\lambda^{\alpha}(\s)$, $\bl^{\a}(\s)$ etc, which 
can also be identified as components of $(1\pm i\Omega)S^2$, 
then have the following mode expansions:
\bea \la{st0}
\l^{\a}(\s) &=& \frac{1}{\a}
\sum_{-\infty}^{\infty}R_n^{\a}e^{in\s/|\a|}, 
\qquad \tilde{\l}^{\a}(\s) ~~~=~~~ \frac{1}{\a}
\sum_{-\infty}^{\infty}R_n^{\a}e^{-in\s/|\a|}, \\
\l^{\ad}(\s) &=& \frac{1}{\a}
\sum_{-\infty}^{\infty}R_n^{\ad}e^{in\s/|\a|},
\qquad \tilde{\l}^{\ad}(\s) ~~~=~~~ \frac{1}{\a}
\sum_{-\infty}^{\infty}R_n^{\ad}e^{-in\s/|\a|},\\
\bl^{\a}(\s) &=& \frac{1}{|\a|}
\sum_{-\infty}^{\infty}\bar{R}_n^{\a}e^{in\s/|\a|},
\qquad \tilde{\bl}^{\a}(\s) ~~~=~~~ \frac{1}{|\a|}
\sum_{-\infty}^{\infty}\bar{R}_n^{\a}e^{-in\s/|\a|},\\ \la{stl}
\bl^{\ad}(\s) &=& \frac{1}{|\a|}
\sum_{-\infty}^{\infty}\bar{R}_n^{\ad}e^{in\s/|\a|}, 
\qquad \tilde{\bl}^{\ad}(\s) ~~~=~~~ \frac{1}{|\a|}
\sum_{-\infty}^{\infty}\bar{R}_n^{\ad}e^{-in\s/|\a|}.
\eea
In writing the above form of the mode expansions, we have used the explicit  
representation of $\Omega$ and $\Pi$ matrices given in eqn.\eq{omegapi},
with $R_n$'s defined in terms of fermionic creation and annihilation
operators as:
\bea\label{Rns0}
R_n^{\a} &=& i{\sqrt{\a}}c_n\Bigl(~\l^{\a}_n + 
\rho_n\l^{\a}_{-n}~\Bigr),\\
R_n^{\ad} &=&i{\sqrt{\a}} c_n\Bigl(~\l^{\ad}_n -
\rho_n\l^{\ad}_{-n}~\Bigr),\\
\bar{R}_n^{\a} &=&-i{\sqrt{\a}}c_n\Bigl(~
\bl^{\a}_n - \rho_n\bl^{\a}_{-n}~\Bigr),\\
\bar{R}_n^{\ad} &=& -i{\sqrt{\a}}c_n\Bigl(~
\bl^{\ad}_n +\rho_n\bl^{\ad}_{-n}~\Bigr),\label{Rnsl}
\eea
and
$\l^{\a,\ad}_n = (\l_{1n}-i\l_{2n})^{\a,\ad}$ and
$\bl^{\a,\ad}_n =(\l_{1n}+i\l_{2n})^{\a,\ad}$. We also have 
$R_0^{\a} = \sqrt{\a}\l^{\a} ,R_0^{\ad} = \sqrt{\a}\l^{\ad}, 
\bar{R}_0^{\a}= \sqrt{\a}\bl^{\a}$ and 
$\bar{R}_0^{\ad} =\sqrt{\a} \bl^{\ad}$ for the zero modes.
One can find the commutation relations of $R'$s, 
using the ones for $\l'$s given in eqns. (\ref{fer-com}) and \eq{ocsil}:
\bea
\{R_m^{\a},\bar{R}_n^{\b}\} &=& \a \d_{m+n,0}\d^{\a\b},\\
\{R_m^{\a},R_n^{\b}\} &=&\{\bar{R}_m^{\a},\bar{R}_n^{\b}\}~~=~~0,
\eea
with similar commutation relations for the $R^{\ad}$'s as well. The bosonic and
fermionic mode expansions given 
in eqns.\eq{xt0}, \eq{pt0} and \eq{st0}-\eq{stl},
have been written in a form so
as to match with the ones given in \cite{GreenIII}. With this matching, it is 
possible to determine the exlicit form of the kinematical 
part of the open string
vertex from the Fourier modes of the constraints:
\bea \label{kinb}
&&~~~~~\sum_{r=1}^3\varepsilon_r x_r(\s_r)|E_b\ra ~~=~~ 0,
~~~~~\qquad
\sum_{r=1}^3 p_r(\s_r)|E_b\ra ~~=~~ 0,\\ \label{kinf}
&&\sum_{r=1}^3\varepsilon_r \l_{(r)\a,\ad}(\s_r)|E_f\ra ~~=~~ 0,\qquad
\sum_{r=1}^3\bl_{(r)\a,\ad}(\s_r)|E_f\ra ~~=~~ 0.
\eea
Here, $r$ is the string index and $\varepsilon_r$ is $+1$ for an
incoming string and
$-1$ for an outgoing one. These conditions generalize the constraints given
for the zero mode vertex in eqns.\eq{kin1}-\eq{kin4}, 
to the full open superstring. This part of the vertex has been derived
in sufficient detail in~\cite{Spradlin1}
with further corrections given in~\cite{Pankiewicz},
for the case of closed strings in PP-wave background.

To determine the bosonic part of the kinematical vertex for our case, 
we write down the constraints given in eqn.\eq{kinb}, in terms of their 
Fourier modes as:
\bea\label{fourierB}
\sum_r \sum_{-\infty}^{\infty} \a_r
X^{(r)}_{mn}x_{n(r)} |E_b\ra ~~=~~ 0,\qquad
\sum_r \sum_{-\infty}^{\infty} X^{(r)}_{mn}p_{n(r)} |E_b\ra ~~=~~ 0,
\eea
where the matrix $X^{(r)}_{mn}$, as well as other functions appearing 
below, are defined in the appendix. These 
constraints on the vertex turn out to be slightly different 
for the Neumann and Dirichlet directions. For Neumann directions
we have ($m>0$):
\bea
\label{nfirst}
\Bigl[~\sum_{r=1}^{3} \sum_{n=1}^{\infty} 
\a_r X^{(r)}C^{-1/2}C^{-1/2}_{(r)} (a^{(r)}_n - a^{(r)}_{-n})
+ i\a B \IR ~\Bigr] |E_b\ra &=& 0,\\
\Bigl[~\sum_{r=1}^{3} \sum_{n=1}^{\infty}
X^{(r)}C^{-1/2}C^{1/2}_{(r)} (a^{(r)}_n + a^{(r)}_{-n})
+ B \IP ~\Bigr]|E_b\ra &=& 0,
\eea
and those in the Dirichlet directions are:
\bea
\Bigl[~\sum_{r=1}^{3} \sum_{n=1}^{\infty} 
X^{(r)}C^{1/2}C^{-1/2}_{(r)} 
(a^{(r)}_n + a^{(r)}_{-n})~\Bigr]|E_b\ra &=& 0,\\
\label{dlast}
\Bigl[~\sum_{r=1}^{3} \sum_{n=1}^{\infty}\frac{1}{\a_r} 
 X^{(r)}C^{1/2}C^{1/2}_{(r)} 
(a^{(r)}_n - a^{(r)}_{-n})~\Bigr]|E_b\ra &=& 0.
\eea
Here $a^{\dagger}_n = \frac{\a^{\dagger}_n}{\sqrt{\omega_{n}}}$ and 
$a_n = \frac{\a_n}{\sqrt{\omega_{n}}}$, satisfy 
$[a_n,a^{\dagger}_m] = \delta_{mn}$. 
Hence, the bosonic part of the kinematical vertex satisfying the coordinate
continuity and momentum conservation constraints
given in terms of their Fourier modes in eqns.\eq{nfirst}-\eq{dlast}, is 
given as 
~\cite{Spradlin1,Stefanski}:
\bea \label{bv}
|E_b\ra &\sim& \exp\Bigl\{~\frac{1}{2}
\sum_{r,s=1}^3\sum_{m,n}
a^{\dagger}_{m(r)}\bar{N}^{rs}_{mn}
a^{\dagger}_{n(s)}~\Bigr\} |0\ra_{123},
\eea
where $|0\ra_{123}=|0\ra_1\otimes|0\ra_2\otimes|0\ra_3$ is annihilated by 
$a_n$. In the above equation, $\bar{N}^{rs}_{mn}$  are the bosonic 
Neumann matrices defined in the appendix and the summation is over 
appropriate modes of Neumann and Dirichlet directions\cite{stefanski2}. 

Similarly, the fermionic part of the kinematical vertex can be determined by
writing down the constraints given in eqn.\eq{kinf} in terms of their 
Fourier modes. The cosine and sine Fourier modes of the $\l^{\a}$ 
conditions are:
\bea
\label{fconsf}
\Bigl[~\sum_{r=1}^{3} \sum_{n=1}^{\infty} 
X^{(r)}C^{-1/2} (R^{\a(r)}_n + R^{\a(r)}_{-n})
- \sqrt{2}\a B \Theta^{\a} ~\Bigr] |E_f\ra &=& 0,\\
\label{fcons2}
\Bigl[~\sum_{r=1}^{3} \sum_{n=1}^{\infty}\frac{1}{\a_r} 
X^{(r)}C^{1/2} (R^{\a(r)}_n - R^{\a(r)}_{-n})
~\Bigr] |E_f\ra &=& 0,
\eea
and the ones for $\bl^{\a}$ conditions are:
\bea
\Bigl[~\sum_{r=1}^{3} \sum_{n=1}^{\infty} 
X^{(r)}C^{-1/2} (\bar{R}^{\a(r)}_n + \bar{R}^{\a(r)}_{-n})
+ \frac{1}{\sqrt{2}}B \bar{\L}^{\a} ~\Bigr] |E_f\ra &=& 0,\\
\Bigl[~\sum_{r=1}^{3} \sum_{n=1}^{\infty}\frac{1}{\a_r} 
\label{fconsl}
X^{(r)}C^{1/2} (\bar{R}^{\a(r)}_n - \bar{R}^{\a(r)}_{-n})
~\Bigr] |E_f\ra &=& 0.
\eea
There will be similar conditions for the Fourier modes of 
$\l^{\ad}$ and  $\bl^{\ad}$ as well.

Hence, the fermionic part of the kinematical 
vertex satisfying the constraints given in terms of their Fourier modes 
in eqns.\eq{fconsf}-\eq{fconsl} (and their dotted counterparts), 
turns out to be~\cite{GreenIII,Spradlin1,Stefanski}:
\bea 
\label{fv}
&& |E_f\ra \sim
\exp \Bigl[~\sum_{r,s=1}^3\sum_{m,n=1}^{\infty}
\Bigl(~\l^{\a}_{-m(r)}~Q^{rs}_{mn\a\b}~
\bar{\l}^{\b}_{-n(s)} ~+~ 
\l^{\ad}_{-m(r)}~Q^{rs}_{mn\ad\bd}~
\bar{\l}^{\ad}_{-n(s)}~\Bigr) \xx
&&\qquad \qquad \qquad-\sum_{r=1}^3\sum_{m=1}^{\infty}
\Bigl( ~\l^{\a}_{-m(r)}~Q^r_{m\a\b}~\bar{\Theta}^{\a}~+~ 
\l^{\ad}_{-m(r)}~Q^r_{m\ad\bd}~\bar{\Theta}^{\ad}~\Bigr)~\Bigr]
|E_f^0\ra,
\eea
where $\bar{\Theta}\equiv\frac{1}{\a_3}(\bar{\l}_{0(1)}-\bar{\l}_{0(2)})$ 
and $|E_f^0\ra$ is the 
zero-mode part of the fermionic vertex given in eqn.\eq{zeroferV}. The
matrices $Q^{rs}_{mn}$ and $Q^r_m$ are the fermionic Neumann matrices. Their
relation to the bosonic Neumann matrices are given in the appendix. 

One can also see the structure of the fermionic part of the kinematic vertex 
written above in eqn.(\ref{fv}), by combining the mode expansions
given in eqns.(\ref{st0})-(\ref{stl}) and (\ref{Rns0})-(\ref{Rnsl}), 
and  comparing with the ones in eqns. (2.6) and
(2.7) of \cite{Pankiewicz}. Alternatively, for example, the mode expansion 
for the quantity $(1-i\Omega)S_2$ (as given in component forms in 
(\ref{st0})-(\ref{stl})) can be written as: 
\bea
(1-i\Omega)S_2(\s)&=& (1-i\Omega)\Bigl[~S_0 ~+~ i\sum_{n = 1}
~\Bigl\{ ~\theta_n \cos \frac{n\s}{|\a|}~ + ~i
\theta_{-n} \sin \frac{n\s}{|\a|}~\Bigr\}~\Bigr],\label{modeS2}
\eea
where,
\bea
\theta_n &=& i\sqrt{\a}
c_n\Bigl[~(1-\rho_n\Pi)S_n + (1+\rho_n\Pi)S_{-n}\Bigr].
\eea
Since these expansions are identical to the ones given in 
~\cite{Pankiewicz}, our Fermionic Neumann matrices are also
identical. In appendix we write their expressions in component forms as
well. We also note that due to a specific choice of basis, matrix $\Omega$
does not appear explicitly in our modes $\theta_n$, as well as in 
kinematic vertex.

Having discussed the kinematical part of the vertex $|V_{kin}\ra$, we go on to 
determine the prefactors, that multiply the vertex and the dynamical 
generators. 
First, we form the following combination of dynamical supercharges given
in eqn.\eq{qminus}:
\bea \la{sup}
&&\sqrt{2p^+}Q^{-} = \frac{1}{2\pi \a^\prime p^+}
\int_{0}^{2\pi \a^\prime |p^+|} d\sigma 
\Bigl[~\Bigl\{\p_{\tau}X^r\gamma^r+ \p_{\s}X^{r'}\gamma^{r'} 
+ m X_r\gamma^r\Omega \Pi \Bigr\}(S^2 + \Omega^T S^1)\xx 
&&~~~~~~~~~~~~~~~~~~~~~~~~~~~~~~~~~
+\Bigl\{\p_{\tau}X^{r'}\gamma^{r'}
+ \p_{\s}X^{r}\gamma^{r} -m X_{r'}\gamma^{r'}\Omega \Pi\Bigr\}
(S^2 - \Omega^T S^1)~ \Bigr].
\eea
Similar combinations of supercharges have been considered in~\cite{GreenI},
for the case of open strings in $D=10$ which in the present context
will correspond to a $D9$-brane in flat space. 
Using the form of  matrices $\Omega$ and $\Pi$ given in eqn.\eq{omegapi}, 
we can now write down the various components of dynamical supercharges 
given in eqn.\eq{sup} as:
\begin{eqnarray} \la{aQ}
&&Q^{-}_{\a} = \int d\s~\frac {1}{\a}\Bigl[~\Bigl\{~P_{\a\ad} 
\l^+_{\ad} + P^+(\ep \l^+)_{\a}+\p_{\s}\tilde{X}^-\bl^+_{\a}
-im(~X_{\a\ad} \l^+_{\ad} -  X^+(\ep \l^+)_{\a})~\Bigr\} \xx
&&~~~~~~~~~~~~+\Bigl\{\tilde{P}^-(\bl^-)_{\a} 
+\p_{\s}X_{\a\ad}\l^-_{\ad} + \p_{\s}X^+(\ep \l^-)_{\a}
-im\tilde{X}^-\bl^-_{\a}~\Bigr\}\Bigr],\\
&&\bar{Q}^{-}_{\a} = \int d\s~\frac {1}{\a}\Bigl[~\Bigl\{~\bp_{\a\ad} 
\bl^+_{\ad} + P^-(\ep \bl^+)_{\a}+\p_{\s}\tilde{X}^+\l^+_{\a}
+im(~\bar{X}_{\a\ad} \bl^+_{\ad} -  X^-(\ep \bl^+)_{\a})~\Bigr\}\xx
&&~~~~~~~~~~~~+\Bigl\{~\tilde{P}^+( \l^-)_{\a} +
\p_{\s}\bar{X}_{\a\ad}\bl^-_{\ad} +\p_{\s}X^-(\ep \bl^-)_{\a}
+im\tilde{X}^+\l^-_{\a}~\Bigr\}\Bigr],\\
&&Q^{-}_{\ad} = \int d\s~\frac {1}{\a}\Bigl[~\Bigl\{~P_{\ad\a} 
\l^+_{\a} + P^-(\ep \l^+)_{\ad}-\p_{\s}\tilde{X}^-\bl^+_{\ad}
+im(~X_{\ad\a} \l^+_{\a} -  X^-(\ep \l^+)_{\ad})~\Bigr\}\xx
&&~~~~~~~~~~~~+\Bigl\{~-\tilde{P}^+( \bl^-)_{\ad} 
+\p_{\s}X_{\ad\a}\l^-_{\a} + \p_{\s}X^+(\ep \l^-)_{\ad}
-im\tilde{X}^-\bl^-_{\ad}~\Bigr\}\Bigr],\\ \la{Qlast}
&&\bar{Q}^{-}_{\ad} = \int d\s~\frac {1}{\a}\Bigl[~\Bigl\{~\bp_{\ad\a} 
\bl^+_{\a} + P^+(\ep \bl^+)_{\ad}-\p_{\s}\tilde{X}^+\l^+_{\ad}
-im(\bar{X}_{\ad\a} \bl^+_{\a} -  X^+(\ep \bl^+)_{\ad})~\Bigr\}\xx
&&~~~~~~~~~~~~+\Bigl\{~ -\tilde{P}^-( \l^-)_{\ad} 
+\p_{\s}\bar{X}_{\ad\a}\bl^-_{\a} + \p_{\s}X^-(\ep \bl^-)_{\ad}
+im\tilde{X}^+\l^-_{\ad}~\Bigr\}\Bigr],
\end{eqnarray}
with the definitions $\l^\pm \equiv \l (\s) \pm \tilde{\l} (\s) =
\l(\s) \pm \l(-\s)$.

For stringy generalization of operators $\IP$ and $\L$ appearing in the last 
section one needs to take care of the singularity at $\s = \pi\a_1$ by 
defining~\cite{GreenIII, Spradlin1} :
\bea \la{def}
P\V &=& \lim_{\sigma \to \pi \a_{(1)}}
-2 \pi \sqrt{-\a} (\pi \a_{(1)} - \sigma)^{1/2}
(~P_{(1)}(\sigma) + P_{(1)}(-\sigma)~)\V,\cr
\p X \V &=& \lim_{\sigma \to \pi \a_{(1)}}
-2 \pi \sqrt{-\a} (\pi \alpha_{(1)} - \sigma)^{1/2}
(~\p_\sigma X_{(1)}(\sigma) + \p_\sigma X_{(1)}(-\sigma)~)\V,\cr
\bL\V &=& \lim_{\sigma \to \pi \a_{(1)}}
-2 \pi \sqrt{-\a} (\pi \a_{(1)} - \sigma)^{1/2}
(~\bl_{(1)}(\sigma) + \bl_{(1)}(-\sigma)~)\V.
\eea
with similar definitions for $\bar{P},\bar{X},X^{\pm},\tilde{X}^{\pm}$. Also,
the same thing holds for the second and the third strings.
The commutation relations involving $\Qa's$ then have identical forms as given
in \eq{qadcomm}-\eq{qdlast}. However, in the present case, there are additional
relations which are non-zero and useful in determining the prefactors:
\bea \la{Qacomm}
\sum_{r=1}^3~ \Bigl\{~\Qa~,\L_{\b}\Bigr\}~&=&~ \p_\s \tilde{X}^- ~\dab,\\
\sum_{r=1}^3~ \Bigl\{~\Qab~,~\bL_{\b}~\Bigr\}~&=&~\p_\s\tilde{X}^+\dab,
\eea
Similarly, for the $\Qad$'s, we have:
\bea \la{Qadcomm}
\sum_{r=1}^3~ \Bigl\{~\Qad~,~\L_{\bd}~\Bigr\}~&=&~-\p_\s\tilde{X}^-\dabd,\\
\la{Qdlast}
\sum_{r=1}^3~ \Bigl\{~\Qabd~,~\Lbd~\Bigr\}~&=&~-\p_\s\tilde{X}^+\dabd.
\eea
Using $X~\V~=~0$ and $\Lambda~\V~=~0$ as in~\cite{Spradlin1},
the state $\V$ can be shown to satisfy:
\bea \la{ppv}
\sum_{r=1}^3~\Qa\V~&=&~-\frac{1}{\a}\p_{\s}\tilde{X}^- \La\V,\\
\sum_{r=1}^3~\Qad\V~&=&~\frac{1}{\a}\p_{\s}\tilde{X}^-\Lad \V,\\
\sum_{r=1}^3~\Qab\V~&=&~-\frac{1}{\a}\Bigl[P_{\a\ad}\Lad
+P^-(\ep\bL)_{\a}\Bigr]\V,\\
\la{qonv}
\sum_{r=1}^3~\Qabd\V~&=&~-\frac{1}{\a}\Bigl[P_{\ad\a}\La 
+P^+(\ep\bL)_{\ad}\Bigr]\V.   
\eea
In writing the action of the dynamical 
supercharges on the vertex, we have also made use of the fact that
$\p_{\s}X^r \V = 0 = P^{r'}\V$, where $r$ and $r'$ are Neumann and 
Dirichlet directions respectively. These relations can be explicitly
verifed in the oscillator basis representation
~\cite{Spradlin2,Pankiewicz,Stefanski}, keeping in mind that the mode
expansions for string coordinates in the Neumann directions are given in 
terms of cosines and for Dirichlet directions in terms of sines. 
      
Now, we can repeat the exercise of section-3 to determine 
the prefactors for the 
superstring vertex, by demanding that the PP-wave superalgebra 
in eqns.\eq{pp1h}-\eq{pp2} is satisfied. For $Q_{a}$'s the form 
of ansatz is same
as given in eqns.\eq{dynqminus}-\eq{dynqbarminus} and for the interaction 
Hamiltonian we have,
\bea \la{prenonzero}
&&|~H~\ra = \Bigl[~P_{\a\ad}~f^{\a\ad}~+~
\bar{P}_{\a\ad }~\bif^{\a\ad}~
+~P^+~f^-\xx
&&~~~~~~~~~~~+~P^-~f^+~+~\p_{\s}\xmt~\tilde{f}^+~+
\p_{\s}\xpt~\tilde{f}^-~\Bigr]~\V, 
\eea
wherein, the quantities $f$'s  are again the functions of   
$\La,\Lad$ which are generalized to include all the string 
non-zero 
modes, with the definitions as in eqn.\eq{def}. Similarly, $P^{\pm}$,
$P_{\a\ad}$ etc. are stringy generalizations of $\IP$'s 
appearing in section-3, and contain all the non-zero creation 
operators. These quantities can be explicitly realized
in the oscillator basis as well. We also note that, unlike in 
flat space, there
may be some $\mu$ dependent normalizations in the above ansatz for
the interaction vertex, corresponding to the difference in functional and
oscillator basis expressions~\cite{Pankiewicz}. However, in the 
following analysis, we ignore these normalizations. 

Using the supercharges given in eqns.\eq{aQ}-\eq{Qlast} and the 
relations \eq{Qacomm}-\eq{qonv} as well as \eq{qacomm}-\eq{qdlast} in the 
PP-wave superalgebra in eqns.\eq{pp1h}-\eq{pp2},
we get several consistency conditions for the unknown functions $f$ 
appearing in 
eqn.\eq{prenonzero}. Most of these conditions are identical to those given in 
eqns.\eq{ffirst}-\eq{flast}. However,
the functions $f$ need to satisfy certain additional conditions. First, the
relations given in eqns.\eq{pp1qm} and \eq{pp1qb} imply the following 
conditions for 
$f_{\a}$ and $\bar{f}_{\a}$:
\bea \la{addf}
f_{\b}\La~+~f_{\a}\Lb&=&0,\\
\frac{\p \bar{f}_{\b}}{\p\Lc} \dac~+~\frac{\p \bar{f}_{\a}}{\p\Lc}
\dbc&=&0.
\eea
Equations \eq{pp2} and \eq{aabard} lead to further restrictions on 
$f_{\bd},f_{\a}$, and $\bar{f}_{\bd}$ as given below:  
\bea
f_{\bd}\La~+~f_{\a}\Lbd&=&0,\\
\bar{f}_{\bd}\La&=&0,\\
\frac{\p f_{\a}}{\p\Lcd}\dbcd&=&0.
\eea
Equation \eq{abarad} gives the following consistency
conditions for $f_{\bd}$ and $\bar{f}_{\a}$:
\bea
\frac{\p f_{\bd}}{\p\Lc}\dac&=~~0,\qquad
\bar{f_{\a}}\Lbd&=~~0.
\eea
Finally, eqn.\eq{abarabard} results in:  
\bea \la{addl}
\frac{\p \bar{f}_{\bd}}{\p\Lc}\dac-\frac{\p \bar{f}_{\a}}{\p\Lcd} \dbcd&=&0.
\eea
We note that the solutions given in eqns.\eq{sol1}-\eq{sol4}, also satisfy
the additional conditions given in the above equations \eq{addf}-\eq{addl} for
the superstring case when proper stringy generalizations 
of various operators are used.
The unknown functions $\tilde{f}^+$ and $\tilde{f}^-$, appearing in the ansatz
for the interaction Hamiltonian given in eqn.\eq{prenonzero} can now 
be determined
by using the algebra in eqn.\eq{pp1h}, which leads to the following equations:
\be \la{tilde}
2\tilde{f}^+ \dab + \frac{1}{\a}\bar{f}_{\b}\La =0,
\ee
and,
\be
2\tilde{f}^- \dab - \frac{\p f_{\a}}{\p\Lc}\dbc=0. 
\ee
Using the solutions for $\bar{f}_{\b}$ and $f_{\a}$ given in equations 
\eq{sol1}-\eq{sol4}, in the above equation, we get (after taking into account
the stringy generalization for $\Lambda$, $\bL$ etc. 
given in eqn.(\ref{def})): 
\be \la{sol5}
\tilde{f}^- = -1,\qquad {\rm and} \qquad \tilde{f}^+ = 
\frac{1}{4\a^2}\eab\La\Lb\eabd\Lad\Lbd.
\ee
The solutions given in equations \eq{sol1}-\eq{sol4}, \eq{hsol1}-\eq{hsol4}
 and \eq{sol5} then 
determine the prefactors, and hence the complete cubic superstring vertex and 
dynamical supercharges for an open string ending on a $D7$-brane.

\section{Conclusions}

In this paper, we have discussed the construction of cubic interaction vertex 
for an open 
string ending on a $D7$-brane in PP-wave background, using Green-Schwarz 
light-cone string 
field theory formalism. We explicitly determined the prefactors 
for the 
cubic interaction vertex and the dynamical supercharges at the level of zero
modes and then generalized it to include all the stringy excitations.
This was
achieved by writing down all the symmetry generators in terms of 
the $D7$-brane 
symmetry structure $\basis$. It is interesting to note that  
prefactors for the 
interaction vertex resemble those of the flat space~\cite{GreenIII}. 
It will  be nice to give an oscillator basis realization of the 
prefactors appearing in the superstring vertex as given for the
flat space case in~\cite{GreenIII}. Such expression for the closed
strings in PP-wave background have already been given in 
~\cite{Spradlin2,Pankiewicz,Stefanski}.
 
The proposal of~\cite{Shiraz} for closed strings, amounts to calculating  
cubic string interaction amplitudes in the light-cone string field approach,
and relating them to the three-point correlation functions on the 
super Yang-Mills
side. In the present case, having
found the cubic open string interaction vertex,  
it would be interesting to 
calculate the dual field theory correlation functions and three-point functions
using the determinant operators given by~\cite{Vijay}, or in terms of
defect CFT correlation functions. This would provide a 
check of the BMN 
duality, for open strings even at the level of  interactions. 
It would also be worthwhile to address the
question of the existence of higher order open string interactions in PP-wave
background, as has been discussed by~\cite{Huang} for the case of 
closed strings.

Finally, although our construction of the cubic vertex was 
for an open string on a
$D7$-brane in PP-wave background, similar analysis should hold for the case
of open string interactions on other $D3$ and $D5$ branes as well. We hope to
get back to these issues in future.

\vskip 1cm
\section*{Acknowledgments}
We are thankful to Ari Pankiewicz for pointing out to us
the relationship between functional and oscillator basis expressions
and for various coments on the draft. We would also like to thank Sanjay 
for useful discussions. 
\appendix

\section{Matrices appearing in the kinematic vertex}
\setcounter{section}{1}
\renewcommand{\thesection}{\Alph{section}}
\setcounter{equation}{0}

\begin{enumerate}
\item
The Matrix $M^{rs}$ appearing in the bosonic kinematic vertex in 
eqn.\eq{zerobosV} is given in~\cite{Spradlin1}:
\bea
M^{rs} &=& \left(\matrix{
\beta + 1 &-\sqrt{-\beta(1+\beta)} &-\sqrt{-\beta} \cr
-\sqrt{-\beta(1+\beta)} &-\beta &-\sqrt{1+\beta} \cr
-\sqrt{-\beta} &-\sqrt{1+\beta} & 0
}\right)
\eea
where $\b =  \a_{(1)}/\a_{(3)}$.
\item
Bosonic Neumann matrices appearing in the exponential of the kinematical
part of the superstring vertex are defined as~\cite{Spradlin1,Pankiewicz}:
\bea
\label{mn} 
&&\bar{N}^{rs}_{mn} =\d^{rs}\d_{mn}
-2\sqrt{\frac{\w_{m(r)}\w_{n(s)}}{mn}}\left(A^{(r)\,T}\G^{-1}A^{(s)}
\right)_{mn}\,,\\
\label{m0} 
&& \bar{N}^{rs}_{m0}  =
-\sqrt{2\m\a_s\w_{m(r)}}\e^{st}\a_t\bar{N}^r_m\,,\qquad s\in\{1,2\}\,,\\
\label{00a} 
&&\bar{N}^{rs}_{00}=
(1-4\m\a K)\left(\d^{rs}+\frac{\sqrt{\a_r\a_s}}{\a_3}\right)
\,,\qquad  r,s\in\{1,2\}\,,\\
\label{00b}
&&\bar{N}^{r3}_{00}  =
\d^{r3}-\sqrt{-\frac{\a_r}{\a_3}}\,,\qquad
r\in\{1,2\}\,.
\eea
where $\a = a_{(1)}\a_{(2)}\a_{(3)}$ and the matrices 
$A^{(r)},\G, \bar{N}^r_m$ and K are defined below.
\bea
\G & = &\sum_{r=1}^3 A^{(r)}U_{(r)}A^{(r)T},\\
U_{(r)} & = & C^{-1}\bigl(C_{(r)}-\m\a_r\bigr),\qquad
C_{mn}  =  m\d_{mn},\\
\bigl(C_{(r)}\bigr)_{mn} & = & \omega_{m(r)}\d_{mn} ~~=~~\sqrt{n^2 +
(\a_{(r)}\mu)^2} \d_{mn},\\
\bar{N}^r & = & -C^{-1/2}A^{(r)\,T}\G^{-1}B,\qquad
K = -\frac{1}{4}B^T\G^{-1}B.
\eea
where the matrices $A^{(r)}$ and $B$ for $m,n > 0$ can be obtained from 
the Fourier transforming the kinematical constraints in 
eqn.\eq{kin1}-\eq{kin4}:
\bea
A^{(1)}_{mn} & = &  (-1)^n\frac{2\sqrt{mn}\b}{\pi}\frac{\sin m\pi\b}
{m^2\b^2-n^2},\\
A^{(2)}_{mn} & =& \frac{2\sqrt{mn}(\b+1)}{\pi}\frac{\sin m\pi\b}
{m^2(\b+1)^2-n^2},\\
A^{(3)}_{mn} & =& \d_{mn},\\
B_m & = & -\frac{2}{\pi}\frac{\a_3}{\a_1\a_2}m^{-3/2}\sin m\pi\b,
\eea
with $\b=\a_{(1)}/\a_{(3)}$.
Below we define the matrices $X^{(r)}$ appearing in 
Fourier expansion of 
constraints in eqn.\eq{fourierB}. 
For $r=1,2$, we have~\cite{Spradlin1}:
\bea
X^{(r)}_{mn} &=&
(C^{1/2} A^{(r)} C^{-1/2})_{mn}
\qquad\qquad\qquad{\rm if}~m,n>0,\xx
&=& {\alpha_{(3)} \over \alpha_{(r)}} 
(C^{-1/2} A^{(r)} C^{1/2})_{-m,-n},
\qquad{\rm if}~m,n<0,\xx
&=& -{1 \over \sqrt{2}} \epsilon^{rs} \alpha_{(s)} (C^{1/2} B)_m\qquad
\qquad{\rm if}~n=0~{\rm and}~m>0,\xx
&=&1~~\qquad\qquad\qquad\qquad\qquad\qquad{\rm if}~m=n=0,\xx
&=&0~~\qquad\qquad\qquad\qquad\qquad\qquad{\rm otherwise}.
\eea
with $X^{(3)}_{mn} = \delta_{mn}$.
\item
Neumann matrices with negative indices are related to the ones with positive
indices as:
$\bar{N}^{rs}_{-m,-n}$. They are related to $\bar{N}^{rs}_{mn}$ via 
~\cite{Spradlin1,Pankiewicz}:
\begin{equation}\label{neg}
\bar{N}^{rs}_{-m,-n}=-\left(U_{(r)}\bar{N}^{rs}U_{(s)}\right)_{mn}\,,
\qquad m,\,n>0\,.
\end{equation}
Further, the following factorization theorem would be useful in verifying that
the kinematical vertex 
satisfies
the kinematic constraints given in eqn.\eq{kinb} and \eq{kinf}
~\cite{Schwarz1,Pankiewicz}:
\bea 
\bar{N}^{rs}_{mn} & = & -(1-4\m\a K)^{-1}\frac{\a}{\a_r\omega_{n(s)}+
\a_s\omega_{m(r)}}
\left[U_{(r)}^{-1}C_{(r)}^{1/2}C\bar{N}^r\right]_m
\left[U_{(s)}^{-1}C_{(s)}^{1/2}C\bar{N}^s\right]_n.
\eea
\item
The fermionic Neumann matrices are related to the bosonic ones 
as~\cite{Pankiewicz}:
\bea
Q^{rs}_{mn} & = & e(\a_r)\sqrt{\left|\frac{\a_s}{\a_r}\right|}
\bigl[P_{(r)}^{-1}U_{(r)}C^{1/2}\bar{N}^{rs}C^{-1/2}U_{(s)}P_{(s)}^{-1}
\bigr]_{mn},\\
\label{qm}
Q^r_n & = & \frac{e(\a_r)}{\sqrt{|\a_r|}}(1-4\m\a
K)^{-1}(1-2\m\a K(1+\Pi))\bigl[P_{(r)}C_{(r)}^{1/2}C^{1/2}\bar{N}^r\bigr]_n.
\eea
with,
\bea
P_{n(r)} &\equiv& \frac{1-\rho_{n(r)}\Pi}{\sqrt{1-\rho^2_{n(r)}}}. 
\eea
Explicitely in our case:
\bea
Q^{rs\ad\bd}_{mn} & = & \sqrt{\a_r \a_s}
\bigl[P_{(r)}^{-1}]U_{(r)}C^{1/2}\bar{N}^{rs}C^{-1/2}U_{(s)}P_{(s)}^{-1}
\bigr]^{\ad\bd}_{mn},\\
Q^{r\ad\bd}_n & = & i\sqrt{2}\frac{\a}{\sqrt{|\a_r|}}
(1-4\m\a K)^{-1}
\bigl[P_{(r)}C_{(r)}^{1/2}C^{1/2}\bar{N}^r\bigr]^{\ad\bd}_n,
\eea
and\\
\bea
Q^{rs\a\b}_{mn} & = & \sqrt{\a_r \a_s}
\bigl[P_{(r)}U_{(r)}C^{1/2}\bar{N}^{rs}C^{-1/2}U_{(s)}P_{(s)}
\bigr]^{\a\b}_{mn},\\
Q^{r\a\b}_n & = & i\sqrt{2}\frac{\a}{\sqrt{|\a_r|}}
\bigl[P_{(r)}^{-1}C_{(r)}^{1/2}C^{1/2}\bar{N}^r\bigr]^{\a\b}_n,
\eea
where,
\bea
P_{n(r)} &\equiv& 
\frac{1-\rho_{n(r)}}{\sqrt{1-\rho^2_{n(r)}}}. 
\eea
\end{enumerate}


\nc{\np}[3]{Nucl. Phys. {\bf B#1}, #2 (#3)}

\nc{\pl}[3]{Phys. Lett. {\bf B#1}, #2 (#3)}

\nc{\prl}[3]{Phys. Rev. Lett. {\bf #1}, #2 (#3)}

\nc{\prd}[3]{Phys. Rev. {\bf D#1}, #2 (#3)}

\nc{\ap}[3]{Ann. Phys. {\bf #1}, #2 (#3)}

\nc{\prep}[3]{Phys. Rep. {\bf #1}, #2 (#3)}

\nc{\ptp}[3]{Prog. Theor. Phys. {\bf #1}, #2 (#3)}

\nc{\rmp}[3]{Rev. Mod. Phys. {\bf #1}, #2 (#3)}

\nc{\cmp}[3]{Comm. Math. Phys. {\bf #1}, #2 (#3)}

\nc{\mpl}[3]{Mod. Phys. Lett. {\bf #1}, #2 (#3)}

\nc{\cqg}[3]{Class. Quant. Grav. {\bf #1}, #2 (#3)}

\nc{\jhep}[3]{J. High Energy Phys. {\bf #1}, #2 (#3)}

\nc{\hep}[1]{{\tt hep-th/{#1}}}


\end{document}